\RequirePackage{fix-cm}
\documentclass[twocolumn]{svjour3_arxiv}          
\usepackage{times}
\usepackage{xspace}
\usepackage{times}
\usepackage{epsfig}
\usepackage{graphicx}
\usepackage{amsmath, bm}
\usepackage{amssymb}
\usepackage{algorithm}
\usepackage{algorithmic}
\usepackage{mathtools}
\usepackage{array}
\usepackage{multirow}
\usepackage[inline]{enumitem}
\usepackage{subfig}
\usepackage{multirow}
\usepackage[utf8]{inputenc} 
\usepackage[T1]{fontenc}    
\usepackage{url}            
\usepackage{booktabs}       
\usepackage{amsfonts}       
\usepackage{nicefrac}       
\usepackage{microtype}      
\usepackage{hyperref}
\usepackage{color}

\smartqed  
\DeclarePairedDelimiter\norm{\lVert}{\rVert}
\makeatletter
\DeclareRobustCommand\onedot{\futurelet\@let@token\@onedot}
\def\@onedot{\ifx\@let@token.\else.\null\fi\xspace}

\def\etc{etc\onedot}
\def\ie{i.e\onedot}
\def\eg{e.g\onedot}

\def\L{\mathcal{L}}

\def\R{\mathbb{R}}

\def\model{DU2MCE}
\newcommand{\pairs}[2]{\emph{(#1, #2)}}
\newcommand{\lambdas}[3]{$(\lambda_1, \lambda_2, \lambda_3) = (#1, #2, #3)$}

\newcommand{\mat}[1]{\bm{#1}}
\newcommand{\set}[1]{\mathbb{#1}}
\def\mW{\mat{W}}

\def\mUe{\mat{U_e}}
\def\sU{\set{U}}
\def\sS{\set{S}}

\def\vx{\mat{x}}

\journalname{IJCV}

\definecolor{redcol}{rgb}{1, 0, 0}
\definecolor{bluecol}{rgb}{0, 0, 1}

\renewcommand{\paragraph}[1]{\smallskip\noindent{\bf{#1}}}
\newcommand{\quotes}[1]{``#1''} 

\begin{document}

\title{Deep Unified Multimodal Embeddings for Understanding both Content and Users in Social Media Networks
}


\author{Karan Sikka         \and
	Lucas Van Bramer 	\and
	Ajay Divakaran
}


\institute{Karan Sikka and Ajay Divakaran are with SRI International, Princeton, NJ, USA\\
	Email: \{karan.sikka, ajay.divakaran\}@sri.com \\
	Lucas Van Bramer is with Department of Computer Science, Cornell University, Ithaca, NY, USA \\
	Email: ljv32@cornell.edu \\
	Lucas did this work as an intern at SRI International. 
}


\maketitle

\def\algorithmautorefname{Algorithm}
\def\figureautorefname{Figure}
\def\tableautorefname{Table}
\def\equationautorefname{Eq.}
\def\sectionautorefname{Section}
\def\subsectionautorefname{Section}
\def\subsubsectionautorefname{Section}

\begin{abstract}

There has been an explosion of multimodal content generated on social media networks in the last few years, which has
necessitated a deeper understanding of social media content and user behavior.
We present a novel content-independent content-user-reaction model for social multimedia content analysis.
Compared to prior works that generally tackle semantic content understanding and user behavior modeling in isolation, we propose a generalized solution to these problems within a unified framework. 
We embed users, images and text
drawn from open social media in a common multimodal geometric space, using a novel loss function designed to cope with
distant and disparate modalities, and thereby enable seamless three-way retrieval.  
Our model not only outperforms
unimodal embedding based methods on cross-modal retrieval tasks but also shows improvements stemming from
jointly solving the two tasks on Twitter data. 
We also show that the user embeddings learned within our joint multimodal embedding model are better at predicting
user interests compared to those learned with unimodal content on Instagram data.  Our framework thus goes beyond the prior practice of
using explicit leader-follower link information to establish affiliations by extracting implicit content-centric affiliations from
isolated users.  
We provide qualitative results to show that the user clusters emerging from learned embeddings have consistent semantics
and the ability of our model to
discover fine-grained semantics from noisy and unstructured data. 
Our work reveals that social multimodal content is inherently multimodal and possesses a consistent structure because in social networks meaning is created through interactions between users and content.


\end{abstract}


\section{Introduction}

Social networks are today's platform of choice for entities that want to exert influence, where the entities range from
well organized groups such as companies and political organizations to amateur individuals
\cite{serrat2017social,hanna2011we,farwell2014media,anger2011measuring}. All contemporary political and commercial
marketing campaigns make skillful and extensive use of social multimedia platforms to engage in direct messaging to the
public using advertising, opinion-analysis and news-facts \cite{hanna2011we,farwell2014media,lipsman2012power}. On the
geo-political front, extremist organizations have successfully recruited for and planned activities through social media
\cite{farwell2014media}. Development of a computational framework for assessing the success of such efforts presents
challenges that go beyond the state-of-the-art in both understanding the content  and understanding the underlying user
behavior and interaction. Social media content derives meaning through the structures and dynamics of interaction
between the multimedia posting, the user who posted the content, and their followers on the social network, rather than
as a stand-alone item to be understood in isolation. For example, a short clip featuring cowboys from a popular movie
posted by the movie's director would be seen as a promotion of the movie, but the same clip when posted by a fashion
house that sells apparel that is featured in that clip would be seen as promoting that line of apparel (see
\autoref{fig:dataset_2}). The success of such influence has been consistently measured in terms of social connections
such as the number of followers and re-tweet statistics
\cite{serrat2017social,kumar2017army,anger2011measuring,leskovec2006patterns}, which does not take generally take the role of the
posted content into account.  Furthermore,  the content on social media networks is highly unstructured with no
constraints on the range of topics or even grammar, thus making state-of-the-art approaches, based on training with
curated datasets, unfit for understanding such content. The nature of the content in such leader-follower configurations
has been studied only in genre specific fashion through investigations of phenomena such as detecting cyber-bullying
\cite{singh2017toward}, hate speech \cite{ribeiro2018characterizing}, pornographic material \cite{singh2016behavioral},
and \quotes{rabble rousing} through politically persuasive content \cite{siddiquie2015exploiting}. There has also been
considerable work on sentiment detection in social media \cite{you2016cross,agarwal2011sentiment}. Such work mostly
focuses on the literal meaning of the posted content rather than the intent behind it. Recently,
\cite{kruk2019integrating} published initial work on identifying document intent in stand-alone Instagram postings based
on the combination on image and its caption.

On the other hand, there has been extensive work on multimedia content analysis independently of social networks,
covering detection and recognition of objects, scenes, activities, concepts, subjective attributes such as sentiment,
visual question answering and topic discovering
\cite{baltruvsaitis2019multimodal,antol2015vqa,veit2018separating,wang2016cnn,faghri2017vse++,chen2015webly,qian2016multi}.
With recent breakthroughs in deep learning, the accuracy of the methods has gone up dramatically. However,
such analysis has yet to be integrated into a unified social multimedia network analysis framework because of the 
challenges stemming from the lack of structure and grammar in posts, unconstrained topics, and heterogeneous data consisting of multiple
modalities which might be missing content from certain modalities in posts depending on the nature of the social networks.

Prior works have also focused on discovering user behavior by learning low-dimensional embeddings either based on
recommendation based models \cite{zhang2019deep,su2009survey,elkahky2015multi,veit2018separating} or using social
network graphs \cite{perozzi2014deepwalk,grover2016node2vec,hamilton2017representation,wang2017signed}. Despite their
success in recommending items to users or predicting user behavior from the learned embeddings, these works have
generally been restricted to social media networks with limited variability in content.  Moreover, these approaches are
generally only concerned with discovering user interests and not with content understanding in a joint manner. We
believe that despite the \quotes{in-the-wild} nature of social media content, it is quite structured and it should be
possible to exploit the relationship between multimodal content and users to enhance the understanding of each task.  In
other words, our aim is to gain a holistic understanding of the nature and propagation of influence in social networks.
To that end, we both need to advance over state-of-the-art approaches so as to deal with the richness and complexity of
social media, as well as to gain a fine-grained understanding of all
the relationships between the various modalities and users.

\begin{figure*}[htbp] 
	\centering
	\begin{center} 		
		\includegraphics[width=0.7\linewidth]{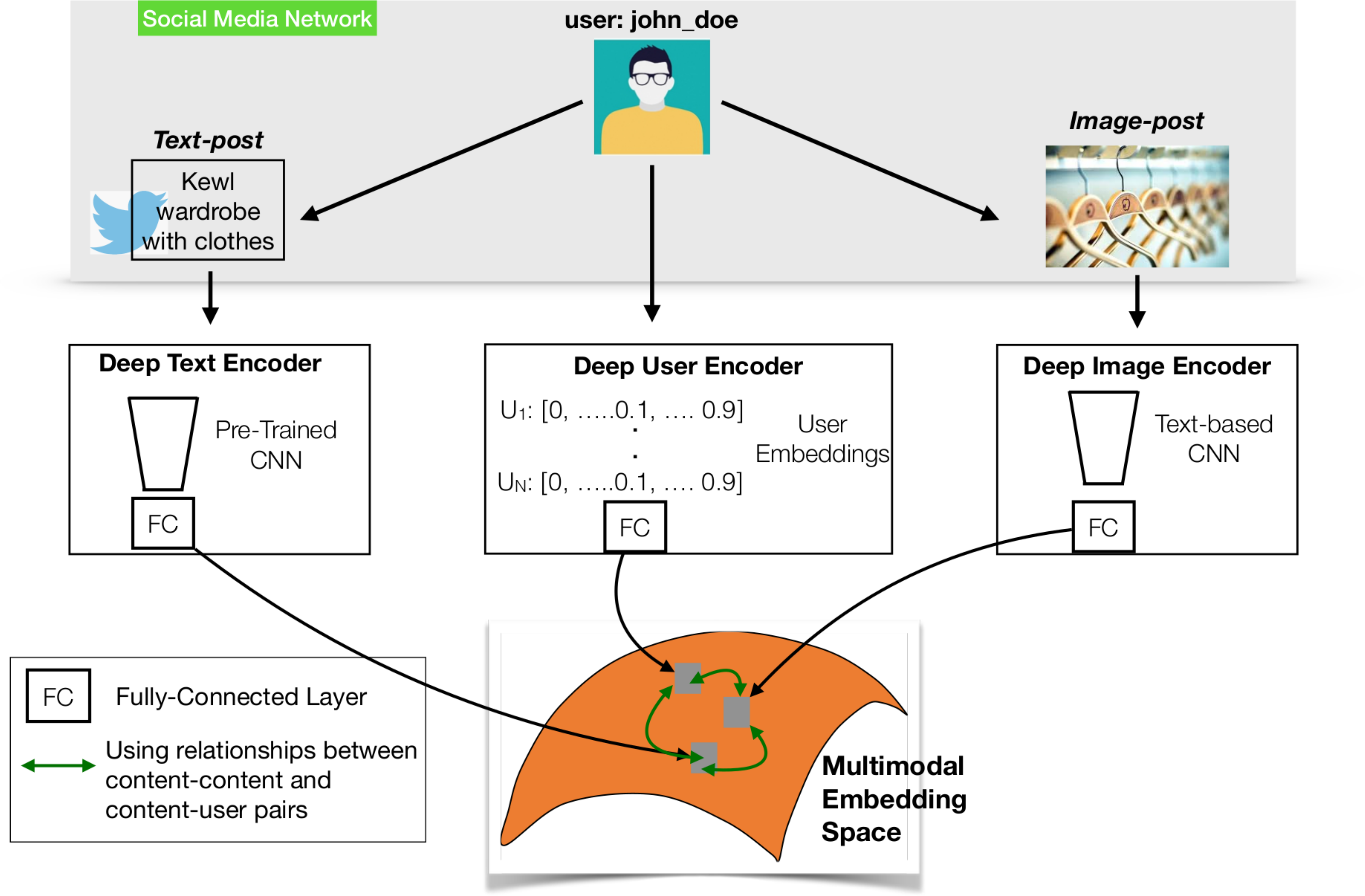}
	\end{center} 
	\caption{This figure shows the block diagram for the proposed Deep Unified User and Multimodal Content Embedding
	Model that learns to embed content (images and text) as well as users in the same geometric space from social
multimedia content. As a result, our model is able to simultaneously tackle content understanding and user behavior
modeling in a unified framework.}
	\label{fig:block} 
\end{figure*}

In this paper, we propose a novel content-independent content-user-reaction model for social multimedia analysis.  We
embed users, images and text drawn from open social multimedia data in a common multimodal geometric space thereby
enabling seamless (bidirectional) three-way retrieval between them. We are thus able to provide a generalized solution to
both the problem of finding user interests and semantic understanding of multimodal content in a unified framework.  We
propose a deep learning based model, which uses modality-specific encoders and embeds their outputs into a common space
(see \autoref{fig:block}). The parameters of this model are learned using a novel loss formulation based on mixture of
pair-wise loss functions designed to tackle the two tasks jointly. We show that our model is able to achieve the best
performance jointly on the task of cross-modal retrieval between content-content pairs and content-user pairs on a
multimodal corpus covering a wide range of topics collected from Twitter.  As a result, our model is able to go beyond
content/modality/genre specific recommendation models developed in previous state-of-the-art methods. Our results also
demonstrate that the cross-modal retrieval between visual and textual content improves when training jointly to
correlate users and content. This shows that user-content information is able to regularize the content understanding
task
due to the strong relationships between users and content.  We then apply the features learned from our joint
multimodal embedding to the task of predicting user interests on data collected from Instagram, and show consistent
improvements compared to embeddings learned from either textual or visual content.  We thus show that our model yields a
general purpose framework for finding user affiliations based on content and without explicitly using social links as used in prior works
\cite{grover2016node2vec,perozzi2014deepwalk}.  
We also show clusters emerging from the learned user embeddings, which seem to be not only grounded in
multimodal content but also possess fine-grained distinctions. Our analysis reveals that social multimedia data is
inherently multimodal and there is an underlying structure due to the social interactions between users and content.

Our specific contributions are listed below:
\begin{enumerate}

	\item We propose a novel approach to simultaneously tackle the problems of modeling user behavior and semantic
		understanding of content from social multimedia data. Our approach embeds both content and users in a
		common geometric space that enables seamless multi-way retrieval between all modalities and users. 

	\item We propose a novel loss function based on mixture of ranking objectives that enables the above mentioned
		multi-way retrieval. Our loss function not only considers the correlations between users-content, as
		done in prior work, but also considers correlations between modalities in content. It has been carefully designed to cope with the distance between the three modalities - image, text and users.  It thus enables a fine-grained component-wise understanding of the contribution of each modality to the influence of a piece of multimodal content.  

	\item We evaluate our approach on two real-world multimodal social media datasets, which capture the diverse topics and intent, levels of structure and grammar, and overall
		unconstrained nature of data shared on contemporary social networks. The datasets are collected
		from Twitter (Multimodal Twitter Dataset) and Instagram (Fashion Instagram Dataset) to account for both text-focused and image-focused platforms. 

	\item We show consistent improvements on the task of cross-modal retrieval between multimodal content pairs and
		content-user pairs. We further show that our proposed loss function enables our model to perform
		zero-shot retrieval \ie reason about modality pairs not seen during training. We also show that the proposed
		framework enhances content understanding when simultaneously learning to discover user interests within of our model.  We thus show
		that user-content interaction in social networks has a strong structure in spite of the completely
		unconstrained user subscription and content posting environment.

	\item We show that the user embeddings learned within our proposed unified framework enable accurate prediction
		of user interests with the Instagram dataset especially when using all the modality pairs for training. 

	\item We show qualitative results that indicate that our framework helps uncover emergent structures (or clusters) in
		social media and their grounding in multimodal content. We find that the emergent clusters are notably consistent in their semantics.

	\item We show qualitative results to highlight the ability of our model to understand content and ground
		fine-grained concepts in multimodal data while learning from noisy social media data. 
		
	\item We also show consistent improvements compared to competitive methods, based on prior
		state-of-the-art, on the combined task of content understanding and modeling user interests evaluated as 
		cross-modal retrieval tasks. 

	\item     Our overall approach enables a holistic understanding of the nature and propagation of influence in
		social networks, through  fine-grained analysis of all the relationships between the data modalities
		including users. It thus helps deal with the richness and complexity of social multimedia data. 

\end{enumerate}

\section{Related Works}
\label{sec:related}

\paragraph{Deep Multimodal Learning:} We have witnessed significant progress in analyzing and understanding multimedia
content in the last decade especially with the recent advances in deep learning
\cite{li2016event,kiros2014unifying,mathews2016senticap,siddiquie2015exploiting,you2016cross,xu2015show,ordonez2011im2text,antol2015vqa}.
Recent works have tackled several problems requiring reasoning over multiple modalities such as visual question
answering (VQA) \cite{antol2015vqa,lu2016hierarchical}, visual caption alignment and generation
\cite{you2016image,kiros2014unifying,faghri2017vse++,anderson2018bottom}, language guided embodied navigational agents
\cite{das2018embodied,anderson2018vision}, visual grounding \cite{datta2019align2ground,anderson2018bottom}, multimodal
sentiment analysis \cite{mathews2016senticap,you2016cross,siddiquie2015exploiting}, and multimodal topic discovery\cite{li2016event,qian2016multi}.  These problems require analyzing data from multiple modalities such as
image, text, and speech and are also referred to as multi-view learning \cite{qian2016multi},
\cite{liu2017hierarchical}, \cite{wang2015deep}.  We refer interested readers to \cite{baltruvsaitis2019multimodal} for
a detailed survey of approaches on multimodal machine learning.  The progress has also been accelerated by the massive
data generation on social media platforms such as Twitter, Instagram, and Facebook, where people not only post text
but also upload images and videos. Despite such rapid progress, deep learning methods typically require large
quantities of carefully curated data for solving the above tasks \cite{zhang2019deep,mahajan2018exploring}. Although
this supervised paradigm is quite popular in deep learning research, it will not be sufficient for understanding
noisy social multimedia content- which is unconstrained in terms of the underlying concepts and topics.  In contrast to
works requiring massive amounts of labeled data, we need to understand multimodal content as well as the interests of the
users posting such content by learning from the weak supervision available from the multimodal posts generated on these
platforms.

\paragraph{Understanding Content and their Semantics from Social Media Data:}
In order to analyze and search the massive social multimedia content, it is
important to develop tools for \emph{higher-level} understanding of content
across different modalities. For example, it is useful to be able to retrieve
content relevant to a sentence or tag-level query across millions of documents
\cite{srihari2000intelligent,shyu2008video,gong2013deep,qian2016multi,ward2013undefined,veit2018separating,zhang2017hashtag}.
This problem is also referred to as extracting semantics from content since it
is generally concerned with understanding content in a form that is intelligible to a
human \cite{sheu2010semantic}. The problem is challenging because of the diversity of topics/concepts, the lack
of training data caused by \emph{one-off} (sparsity) occurrence of much content, the
multimodality of the data and the completely unconstrained in-the-wild
content capture conditions. Prior works can be divided into
supervised and unsupervised methods. \emph{Supervised} approaches use human curated
data for learning machine learning models that can then be used to extract
semantic tags \cite{wang2016cnn}, descriptions \cite{faghri2017vse++} or
subjective attributes (sentiments \cite{borth2013large}, sarcasm
\cite{schifanella2016detecting}, metaphors \cite{shutova2016black}) from
social media content.  The most common practice is to annotate images by
predicting multiple semantic tags by using classifiers trained for
detecting objects, scenes, attributes \etc \cite{gong2013deep,wang2016cnn,shutova2016black,zhou2014learning}. Despite
their popularity, these approaches suffer from two critical drawbacks: (1) the number
of tags is fixed and cannot be extended to new classes without re-training, and
(2) these detectors generally do not take into account the semantic overlap between tags. Although recent works
have tried to tackle these issues through zero-shot learning \cite{fu2017recent,dalton2013zero,bansal2018zero} or 
by modeling the relationships between tags for improving
recognition \cite{gong2013deep,wang2016cnn}, the issues are far from solved. These 
drawbacks limit the applicability of such approaches in analyzing  social multimedia data from multiple platforms at scale.

Image caption generation and alignment are other alternatives to solve the
problem of understanding social multimedia content by describing/matching an
image with a human level sentence
\cite{hossain2019comprehensive,faghri2017vse++,kiros2014unifying,xu2015show,datta2019align2ground}.
This sentence is expected to describe the salient parts of an image such as
scene, attributes, activities and interactions. Over the last three years we
have seen significant progress in image captioning with the advances in vision
and language based deep learning \cite{hossain2019comprehensive}. These methods
generally rely on learning a joint metric space where images and sentences (or
descriptions) are embedded together. Based on the application, the joint space
can be tuned for for cross-modal matching or generating a caption in the image
captioning task or answering a question in the VQA task.  Despite promising
results, current methods suffer from several limitations for downstream tasks, such as cross-modal retrieval, 
requiring
understanding of social multimedia content. First, these methods require a
well-structured training dataset e.g. MSCOCO \cite{lin2014microsoft} where
humans are given a set of clear instructions for writing captions for given
images.  This is not only a time-consuming process but also the feasibility of
obtaining factual and grammatically correct sentences from the data generated
on social media platforms such as Twitter is limited \cite{ordonez2011im2text}.
Another key drawback is that these approaches are not known to work well for
previously novel objects and scenes. Although there has been recent effort
towards targeting captions for unseen objects \cite{anne2016deep}, it is still
an open problem. Our work addresses the problems with collecting large amounts
of supervised data by proposing to learn the underlying semantics in multimodal
data from large-scale multimodal data-- visual and textual
(\autoref{fig:dataset_1}), while also modeling the users posting such content
into account.  Our algorithm is motivated by prior works utilizing joint
multimodal spaces for aligning between multimodal entities \eg images and text
\cite{kiros2014unifying}. However, we also embed users posting such content in
the same multimodal space leading to better performance on cross-modal
retrieval task on social multimedia data.

Recently, several approaches have explored the use of \emph{data-free} methods
for understanding content. These approaches either use unsupervised approaches
for discovering the underlying semantics and/or topics
\cite{chen2013neil,li2016event,qian2016multi} or utilize the supervision
available from the noisy tags for learning predictive models
\cite{weston2014tagspace,denton2015user,veit2018separating,chen2015webly}. For
example, \cite{qian2016multi} use topic modeling to automatically discover
multimodal topics and opinion from disparate news sources and modalities.
\cite{li2016event} propose to mine multimodal concepts relevant to specific
events from collections of images and captions from sources such as news
websites and Twitter.  Works based on using supervision available from noisy
user annotations, referred to as \emph{webly supervised} learning
\cite{chen2015webly}, have focused on problems such as learning to predict
hashtags
\cite{zhang2017hashtag,veit2018separating,denton2015user,weston2014tagspace},
learning pre-trained models or visual features for downstream tasks
\cite{mahajan2018exploring,chen2015webly,joulin2016learning}, learning to
predict visual concepts/descriptions
\cite{ordonez2011im2text,gan2016you,xu2018webly,zhuang2017attend}.   Our work
is closely related to \cite{veit2018separating,denton2015user}, which focus on
personalized hashtag prediction for a user. Specifically, they propose to learn
models for predicting hashtags for an image conditioned on the user
preferences- that can either be learned or obtained from the meta-data.  The
proposed work differs from the prior works in several aspects. First, our
model directly associates images with sentences instead of individual tags,
which leads to a richer semantic grounding to describe fine-grained
concepts and is not restricted by a discrete vocabulary of the hashtags.  For
example, as shown in \autoref{fig:retrieval}, our model is able to utilize the
compositional nature of sentences and learn to differentiate between the
semantics of two closely related compound concepts-- \quotes{healthy food}
and \quotes{unhealthy food}. This enables our model to better tackle the richness of social multimedia content. Second, compared to
\cite{veit2018separating,denton2015user}, which are restricted to learning
associations between multimodal content, the proposed work learns to associate
multimodal content as well as user interests within the same deep learning
based framework.  Another key difference is between the use of Instagram and
Twitter social networks, used for evaluation in our work, which have 
completely different social dynamics in comparison to passive uploading sites
such as Flickr as used in prior works \cite{veit2018separating,denton2015user}.
Platforms such as Flickr serve as repositories for users to upload their
photographs which are mostly related to personal and recreational pictures. On
the other hand, content on Instagram and Twitter is predicated on provoking a
response from the audience and can convey a multitude of intents based on the
specific use of visual and textual modality \cite{kruk2019integrating} (see \autoref{fig:dataset_1} and
\autoref{fig:dataset_2}).  Moreover, the images are quite noisy, diverse and
not restricted to specific genres.  These images are often intimately
associated with textual content and can be understood only in the context of a
certain thread stemming from a hashtag.

\paragraph{User-Content Recommender Models:} 
The proposed model is closely related to prior works on recommendation based
models in which the task is to learn to recommend items to a user based on their
past preferences, usage patterns \etc \cite{zhang2019deep,su2009survey}.  The
models can be broadly divided into collaborative filtering, content based
recommender system, and hybrid models
\cite{su2009survey,zhang2019deep,pazzani2007content,burke2002hybrid,fang2015collaborative}.
Collaborative filtering exploits prior information about user-item interactions
to learn a vector representations for users and items. These
representations capture the attributes of content and users, and are used for
recommendation \cite{sarwar2001item,he2017neural}.  While, content based
recommendation methods directly use the features of items or users to recommend
similar items \cite{pazzani2007content}.  Our work broadly falls into the
category of hybrid recommendation systems \cite{burke2002hybrid}, that combine
both content representation and use of prior knowledge about user-item interaction
(collaborative filtering).

Recent recommendation methods have also used neural networks due to their
advantage in modeling non-linear interactions and power of representation
learning across different modalities
\cite{fang2015collaborative,he2017neural,dziugaite2015neural,zhang2019deep}.
\cite{he2017neural} extend collaborative filtering to Neural Collaborative
Filtering (NCF) by using neural networks to learn latent representations for
users and items based on prior interaction data. They showed that NCF is able
to surpass the performance of non-deep methods by a significant margin.
\cite{dziugaite2015neural} propose an extension of matrix factorization by
using neural networks to replace the inner product operation and use it for
collaborative filtering.  Several methods have also explored the use of
modality-specific deep encoders for learning joint representations for
(heterogeneous) content and users, which can then be used for recommendations
\cite{denton2015user,veit2018separating,elkahky2015multi,lei2016comparative,zhang2017joint,huang2013learning}.
\cite{elkahky2015multi} proposed a cross-domain and multi-view recommendation
system with multiple domain (news, app data, movie/TV usage) using neural
networks. They learn a similarity function between a user and different views
based on their interaction data.  Despite solving the problem on handling data
from different domains, there are noteworthy differences compared to our work.
First, we work with distant domains (or modalities), namely visual and textual,
that are very different from each other in comparison to the text-based domains
used in \cite{elkahky2015multi}.  Second, our loss function considers
correlations not only between users-content, as done in
\cite{elkahky2015multi}, but also within content. Finally,
\cite{elkahky2015multi} uses a single loss function to combine the scores from
different user-domain pairs, while  our loss function achieves it by using a
mixture of objectives which is more efficient at handling missing and unbalanced
data. Similarly, our work also differs from \cite{zhang2017joint}, who also
propose to learn joint representations for different modalities and users based
on product reviews.  However, our work differs from this work in learning to
associate not only user-content interactions but also content-content
interactions.  Our loss function is also different from \cite{zhang2017joint},
who merge the modalities into a single vector prior to optimizing a single
objective function correlating merged content and user.  Finally, as previously
noted we evaluate our model on social networks such as Instagram and Twitter,
whose content is more unconstrained and has a wider range of topics and
semantics as compared to the dataset of product reviews used in \cite{zhang2017joint}. Compared to
most of these prior works, the proposed model aims to simultaneously address
the problems of understanding content and user behavior from social multimedia
data.

\paragraph{Social Network Embeddings:} Prior works have also looked into
embedding users based on the social network graph or structure. A social
network can be modeled as a graph with users as the nodes and edges as the
connection between them such as friends, followers, commenters. These methods
generally utilize the graph or neighborhood structure and embed users nodes
into a low-dimensional embedding space capturing their structural similarities.
Works such as
\cite{grover2016node2vec,wang2016structural,perozzi2014deepwalk,hamilton2017representation}
use the local first or second order neighbourhood of nodes and convert  the
problem of learning node embeddings as an unsupervised representation learning
problem. Similar to other areas, recent works have also used deep learning to
improve the quality of the learned embeddings. For example,
\cite{perozzi2014deepwalk} treat a social network as a set of documents and
propose an algorithm combining random walk and skip-gram model (used in
learning unsupervised word representation). The learned embeddings can then be
used for downstream tasks such as prediction of links between user or
predicting their interests. We refer the interested reader to
\cite{hamilton2017representation} for a detailed survey on these methods. A key
limitation with the idea of learning user representations from social network graphs
is that such information might not always be available from different social
network platforms. In such cases, our approach provides an alternative mechanism to learn user
interests (via embeddings) by relying on their posted content.
Our approach also provides the benefit of being able to discover interests for
isolated users in a network.   We also believe that our approach can be
combined with these approaches to gain from their relative complementary strengths in
discovering user behavior.

\label{sec:approach}

\section{Approach}

In this section we describe our approach, \textbf{Deep Unified User and Multimodal Content Embedding Model} (\model), in
detail. The proposed model meets the requirements of understanding multimodal content as well as discovering users'
affinity to such content (reaction) from open social media data in a common computational framework.  We achieve this by
learning to embed both multimodal content (visual and textual) and users in a common geometric space as shown in
\autoref{fig:block}. \model~represents multimodal content as a semantic vector in the common embedding space.  Each
user in the given snapshot of the social media network is also represented as a vector in the common space.  The learned
vector implicitly captures the interests and behavior/affiliations of that user and can also be used to measure the
reaction of that user to any content. Here, we refer to reaction as a measure of the user's interest in
certain content without any measure of polarity. 
The model learns to embed content and user in an unsupervised manner by utilizing
the correlations between users and their multimodal posts.

We propose a novel end-to-end deep learning based approach that uses an innovative loss function based on multiple
objectives to embed users and multimodal content in a common embedding space. 
We next describe our approach mathematically.

\subsection{Deep Unified User and Multimodal Content Embedding Model (\model)}

\model~embeds users and multimodal content from a social media network is a unified geometric space. We denote the  $j^{th}$ post $\mathcal{P}_j$ as a
triplet $\mathcal{P}_j=(I_j, T_j, U_j)$, where $I_j \in \emptyset_I \cup \mathcal{R}^{M \times N \times 3}$ is the
posted RGB image, $T_j \in \emptyset_T \cup \sS$ is the posted text, and $U_i \in \sU$ is the user who authored the
post. $\sS$ is the set of all strings and $\sU=\{U_k\}_{k=1}^{K}$ is set of all $K$ users in the corpus. We use 
$\emptyset_I$ and $\emptyset_T$ to denote null elements for the case of missing data from image and text modalities
respectively from a post. 
Being able to handle missing multimodal content in a post is a important requirement since users can author posts not
containing either text or images. For example, users generally post more images than text on Instagram as compared to Twitter or Reddit.

We encode multimodal content by using modality-specific deep encoders.  We denote the deep encoder for image and text
modalities as $\phi_I$ and $\phi_{T}$. By the virtue of using a common geometric space for embedding content and users,
we mathematically regard users as another modality such that each user $U_k$ is encoded as a fixed length vector by an
encoder denoted as $\phi_U$\footnote{We shall refer to users as
another modality interchangeably throughout the paper.}. We describe each of these encoders below:

\begin{enumerate}

	\item Image encoder ($\phi_I$)- we use a convolutional neural network
		(CNN), pre-trained for image classification, and remove the last layer and replace it by a fully-connected (FC) layer for projecting visual
		features into the common space. We denote the linear projection layer as $\mW_{I} \in \R^{D_I \times D}$,
		where $D_I$ and $D$ is the dimensionality of the image-features and the common embedding space
	respectively. We denote the output of this encoder for the image from post $\mathcal{P}_j$ as $\vx^I_j = \mW_{I} \phi_I(I_j)$. 

\item Text encoder ($\phi_T$)- we use a CNN based sentence encoder based on word based embeddings \cite{kim2014convolutional} and use a FC for projecting
		features to the common embedding space.  We denote the linear projection layer as $\mW_{T} \in \R^{D_T
		\times D}$, where $D_T$ is the dimensionality of the text-features.  We denote the output of this
		encoder for the text from post $\mathcal{P}_j$ as $\vx^T_j = \mW_{T} \phi_T(T_j)$. 

	\item User encoder ($\phi_U$)- we use an embedding matrix $\mUe \in \R^{K \times D_u}$, whose $i^{th}$ row
		corresponds to the $i^{th}$ user in our corpus and  
	$D_u$ is the
		dimensionality of the embeddings. We use a FC for projecting features to the final embedding
		space. We denote the linear projection layer as $\mW_{U} \in \R^{D_U
		\times D}$. We denote the output of this
		encoder for the user from post $\mathcal{P}_j$ as $\vx^U_j = \mW_{U} \phi_U(U_j)$. 
\end{enumerate}

We learn the parameters of the above encoders and the projection layers by using a ranking based loss function that
enforces co-occurring pairs of content/user to occur closer to each other in the embedding space and non-co-occurring pairs to be farther in the embedding space
\cite{parkhi2015deep}, \cite{kiros2014unifying}. For example, we would expect the embeddings of content related to
fashion and users interested in fashion to be close to each other.  Compared to prior works on cross-modal embeddings
that generally handle data from a pair of modalities \cite{kiros2014unifying} or a user-content pair from unimodal
content
\cite{huang2013learning}, we want to embed content from multiple modalities along with users.

We achieve this by proposing a loss function that uses a mixture of pair-wise objectives that jointly enforce the
multimodal content and the user authoring this content, to be closer in the common embedding space.  Specifically, the
proposed loss optimizes the proposed DCNN architecture so as to push closer co-occurring pairs of multimodal content
(\pairs{image}{text} pairs) as well as 
the content-user pairs (\pairs{text}{user} and \pairs{image}{user} pairs).  We denote the paired
loss function for the \pairs{text}{user} pairs as $\L_{T-U}$, \pairs{image}{text} pairs as $\L_{I-T}$, and
\pairs{image}{user} as $\L_{I-U}$.  The final loss is given as a convex combination of these losses:

\begin{align}
	\L &= \frac{\lambda_1}{N_{T-U}}\L_{T-U} +  \frac{\lambda_2}{N_{I-T}} \L_{I-T}+ \frac{(1 - \lambda_1 -
	\lambda_2)}{N_{I-U}} \L_{I-U} \\
	   & s.t. \quad \lambda_1, \lambda_2  \leq 1
	   \label{eq_loss}
\end{align}

where $\lambda_1$ and $\lambda_2$ are regularization parameters controlling the relative contribution of learning from
different multimodal (or user) pairs. Here $N_{T-U}$, $N_{I-T}$, and $N_{I-U}$ are the number of valid (\emph{text},
\emph{user}), (\emph{image}, \emph{text}), and (\emph{image}, \emph{user}) pairs respectively inside a minibatch. The
regularization parameters can either be set empirically based on a validation set or chosen based on the ratio of the
volume of visual modality and the volume of textual modality. Our work uses these pair-wise losses to
generalize prior works on learning joint representations from heterogeneous sources \cite{zhang2017joint,elkahky2015multi} by not only learning to associate user-content but also modalities within cross-modal content. The
proposed loss function allows us to perform zero-shot retrieval on unseen modality pairs during training and also better
handle missing data, in comparison to prior works because of the mixture based formulation (see
\autoref{sec:sota}).

We compute each of the paired loss functions by using a max-margin based ranking loss formulation that samples an anchor
sample from one modality and a positive and negative sample from another modality.  We describe the formulation of the loss with an example
for computing the loss for a (\emph{text}, \emph{user}) pair from post $\mathcal{P}_j$ \ie $\L_{T-U}(T_j, U_j)$. We first extract text and user features from
post $\mathcal{P}_j$ as $\vx^T_j$ and $\vx^U_j$ respectively. We then sample negative user pairs $\sU_{\prime} =
\{U_n\}_{n=1}^{N} \quad s.t. \quad U_n \neq U_j$ .  We use a cosine similarity function to compute the similarities between text and user pairs as
they are embedded in the same space by the proposed deep encoders. The loss and the similarity metric are computed as: 
 
\begin{align}
	S(\vx^T_j, \vx^U_n) &= \frac{\vx^T_j\vx^U_n}{\norm{\vx^T_j}_2\norm{\vx^U_n}_2} \\
	\L_{U-T}(T_j, U_j) &= \sum_{U_n \in \sU_{\prime}} [0, m - S(\vx^T_j, \vx^U_j) + S(\vx^T_j, \vx^U_n)]_+ 
\end{align}

where $[a]_+ = \max(a,0)\ \forall a$ and $m$ is the margin. 
We sample the negatives ($\sU_{\prime}$) in an online fashion from a sampled
minibatch. We handle the case of missing data from either modality by summing \autoref{eq_loss} over pairs available
from non-missing samples and then normalizing the loss components appropriately. 
The above formulation guides the model to increase the
similarity between pairs from the same post and vice-versa. We use the same formulation and similarity
metric for $\L_{I-T}$ and $\L_{I-U}$.

\section{Experiments}

\begin{figure*}[htbp!] 
	\centering
	\begin{center} 		
		\includegraphics[width=0.6\linewidth]{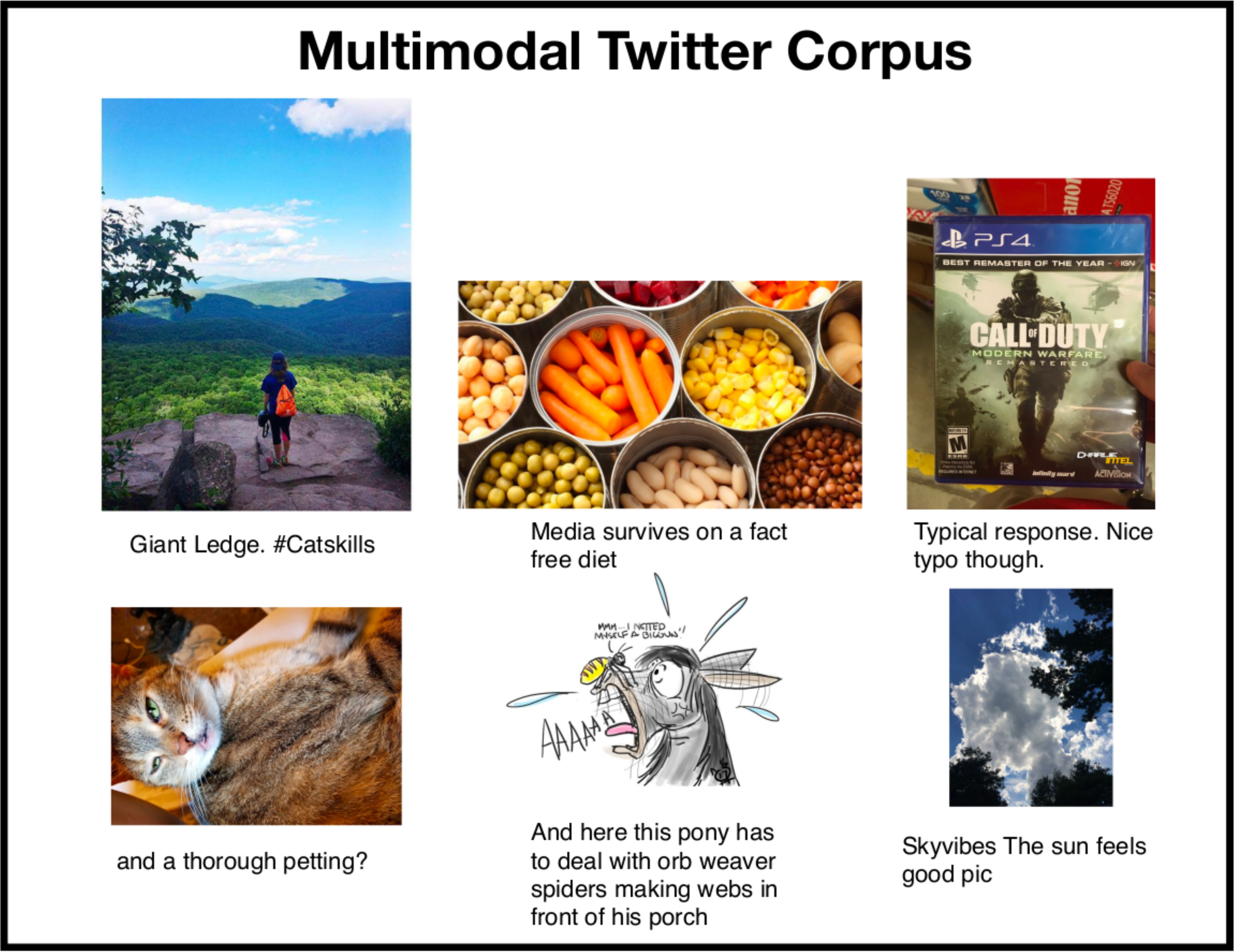}
	\end{center} 
	\caption{This figure shows some images and their corresponding tweets from the proposed Multimodal Twitter Corpus.
		These examples depict the \emph{in-the-wild} nature of social multimedia content. It
also shows how users use both images and captions to convey specific meaning from their posts
\cite{kruk2019integrating}. For example, for the second image in first row, showing some raw food items, it is
difficult to establish that the intent conveyed by the user is not about food but rather about their perception of news media solely
based on the image.}
	\label{fig:dataset_1} 
\end{figure*}

\begin{figure*}[htbp!] 
	\centering
	\begin{center} 		
		\includegraphics[width=0.6\linewidth]{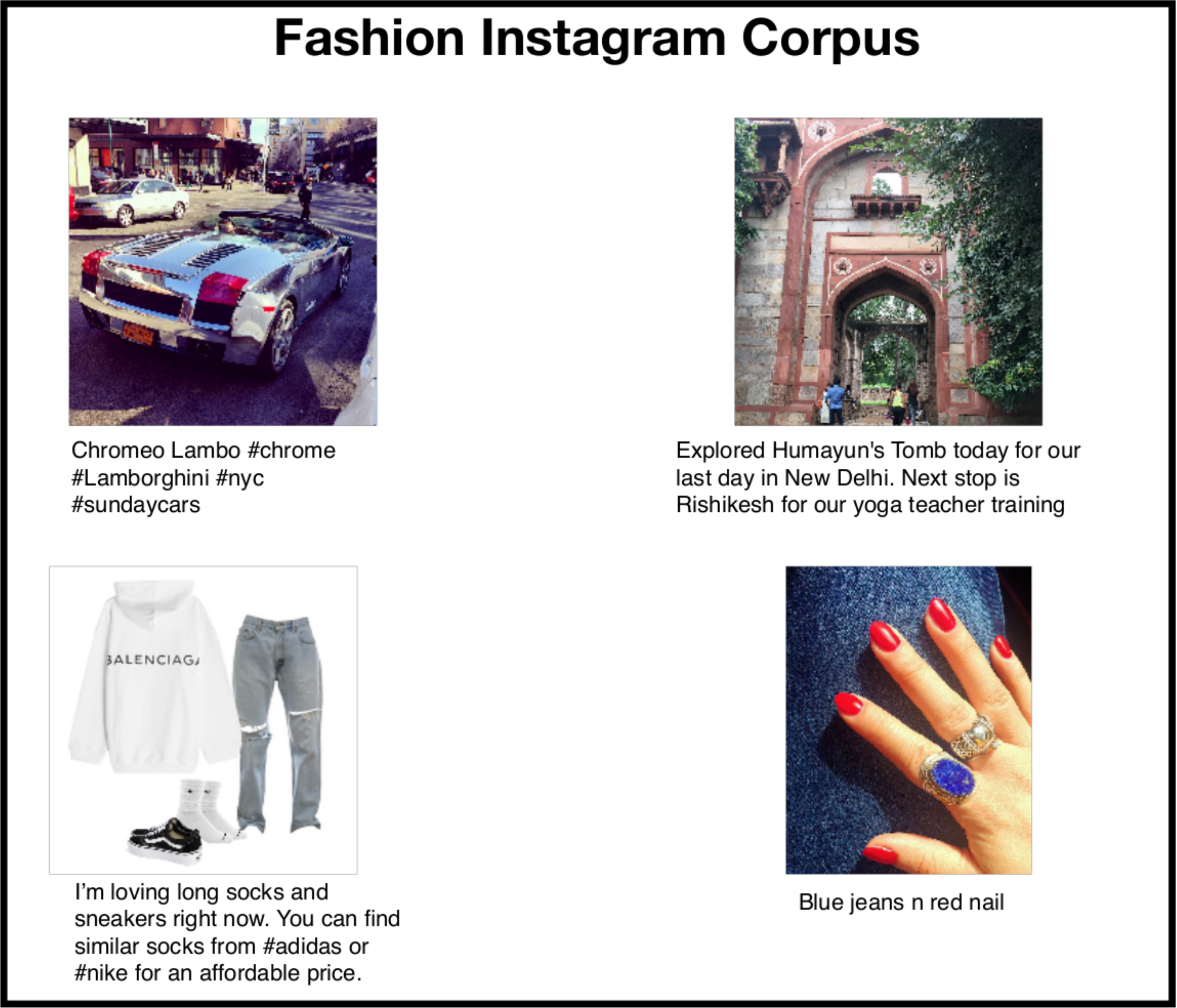}
	\end{center} 

	\caption{This figure shows some posts from the proposed Fashion Instagram Corpus. These images demonstrate
	the strong visual nature of social media posts on Instagram platform as compared to Twitter. Similar to other
multimodal social media networks, users use both captions and images to create specific meaning or intent from their
posts \cite{kruk2019integrating}. For example, despite the two posts on the second row being about fashion, they
convey different intents. The first is about spreading general information about a product while the second is about
showing off or exhibiting personal preferences.} 

	\label{fig:dataset_2} 
\end{figure*}

We now describe the experiments used for performance evaluation. We begin by outlining the datasets used to evaluate our
approach. Since there were no prior public datasets that capture the complexity of multimodal content and users on
current social media networks, we collect two datasets from Twitter (\emph{Multimodal Twitter Corpus}) and Instagram
(\emph{Fashion Instagram Corpus}). We then provide information regarding the tasks and corresponding metrics used for
evaluation. This is followed by information regarding implementation details.  Thereafter, we describe the empirical
results.  In particular, we first evaluate the performance of our model directly on cross-modal retrieval tasks for
content understanding and recommending content to users based on their interests.  We then evaluate the ability of the
user embeddings learned from our model on a downstream task of predicting user interests on the Fashion Instagram
Corpus.  We then provide qualitative results to show the salient (multimodal) topics captured by the user clusters derived
from the learned user embeddings on the Multimodal Twitter Corpus.  We finally compare our model with some competitive
methods and baselines on cross-modal retrieval tasks on the Multimodal Twitter Corpus.

\subsection{Datasets}
\label{sec:datasets}

In order to perform a comprehensive evaluation, the datasets should mirror the challenges associated with unconstrained
and noisy (multimodal) content and user behavior on social media platforms. 
We meet these requirements by using two datasets
from popular social networks for evaluation.   The first dataset, referred to as the Multimodal Twitter Corpus, contains
popular content as guided by the $100$ most popular Twitter hashtags.  The second dataset, referred to as the Fashion
Instagram Corpus, contains content from users who follow some popular fashion, hospitality, and shoe brands on
Instagram. These datasets also allow us to study the performance of our model on network with different content
characteristics \eg Twitter has more textual content than visual content, which is the opposite for Instagram.   
We next describe the methodology for collecting these datasets followed by pre-processing and creation of
train-test splits for evaluating our model. The statistics for these datasets are reported in \autoref{tab:corpus}. 

\begin{table*}[htbp!]
\centering
\begin{tabular}{c|c} \hline
	\multicolumn{2}{c}{Multimodal Twitter Corpus} \\ \hline 
	$\#$ Tweets & $9.45M$ \\ 
	$\#$ Images & $1.26M$ \\
	$\#$ Users  & $39808$ \\ \hline
	\multicolumn{2}{c}{Training Set (Multimodal Embeddings)} \\ \hline 
	$\#$ Tweets & $7.10M$ \\ 
	$\#$ Images & $981K$ \\ \hline 
	\multicolumn{2}{c}{Testing Set (Cross-Modal Retrieval)} \\ \hline 
	$\#$ User-Text pairs & $1.89M$ \\
	$\#$ Image-Text pairs & $5K$ \\ 
	$\#$ Image-User pairs & $251K$ \\ \hline
\end{tabular} \qquad \qquad \qquad
\begin{tabular}{c|c} \hline
	\multicolumn{2}{c}{Fashion Instagram Corpus} \\ \hline
	$\#$ Posts & $796K$ \\ 
	$\#$ Images & $796K$ \\
	$\#$ Captions & $722K$ \\
	$\#$ Users  & $2900$ \\ \hline
	\multicolumn{2}{c}{Predicting User Interests} \\ \hline
	$\#$ Users  & $2877$ \\ 
	$\#$ Users interested in Shoes & $739$ \\ 
	$\#$ Users interested in Fashion & $881$ \\ 
	$\#$ Users interested in Hospitality & $1758$ \\ 
\end{tabular}

\caption{Statistics for the two Multimodal social media Corpora used to evaluate the
	proposed algorithm. The Multimodal Twitter Corpus is collected from Twitter and is used for the cross-modal
	retrieval experiments for evaluating both the tasks of content understanding and discovering user interests.  
	The Fashion Instagram Corpus, collected from Instagram, is used to
	evaluate the task of predicting user interests from the user embeddings learned by our model. We have shown
examples from these datasets in \autoref{fig:dataset_1} and \autoref{fig:dataset_2}.}

\label{tab:corpus}
\end{table*}

\paragraph{Multimodal Twitter Corpus:}
We collected this dataset by first creating a list of $100$ most popular hashtags on Twitter \eg \#love, \#beautiful,
\#me, \#cute, \#nature, \#amazing \etc. These hashtags were then used to crawl an initial list of $\sim 110K$ accounts
from Twitter. We then downloaded the tweets-- text and posted images (if available) from these accounts. Since our aim is
to gauge user behavior from multimodal content only and not social connections, we do not use any user metadata such as location,
followers for learning our model.  Due to the wide variety in the initial hashtags, this dataset contains content
and users interested in diverse topics such as nature, art, international events, fashion, sports, movies \etc.  We have
shown some examples from the dataset in \autoref{fig:dataset_1}. It is evident from these examples that this dataset
contains noisy image-caption pairs, where they seem to be interacting to convey a specific meaning
\cite{kruk2019integrating}.  For example, for the second image in first row in \autoref{fig:dataset_1}, showing some raw
food items, it is difficult to establish that the intent conveyed by the user is not about food but rather about 
news media, solely based on the image. We believe that it is important to work with such examples to 
understand the multitude of user interests on present social media platforms. Prior to pre-processing, we include accounts with more
than $20$ tweets resulting in $\sim40K$ users.  As shown in \autoref{tab:corpus}, the final dataset after pre-processing
has $39808$ users and includes around $9.45M$ posts of which all of the posts have text and only $1.26M$ posts have
images. The relative ratio of different modalities is a function of the social interactions and nature of the social
media networks. For example, platforms such as Instagram and Snapchat are more visual as compared to Twitter or Reddit.
We use this dataset for the cross-modal retrieval experiments.

\paragraph{Fashion Instagram Corpus:}
We collected this dataset by first using a list of $23$ popular brands strongly focused on one of the
three categories- \quotes{shoes}, 
\quotes{fashion}, 
and \quotes{hospitality}. 
We then retrieve a few hundred posts from these brands. We first expanded the user set by retrieving posts
from users who made a comment on these posts. We also added additional data from users who followed these brands but were not
included in the original list. This dataset only includes a subset of users since we are restricted by the Instagram API.
We have shown some samples from the dataset in \autoref{fig:dataset_2}, which clearly shows the visual nature of content
employed by users for
self-expression.
As shown in \autoref{tab:corpus}, the final dataset after pre-processing has $2900$ users and includes around $796K$ posts of which almost all the posts have
images and around $722K$ posts have captions. This is in contrast to the previous dataset where the ratio of images to
text was $1:10$. We use this dataset by employing the user embedding learned by our model for the downstream task of
predicting user interests.

\paragraph{Pre-Processing and Training/Test Splits:}
We focus on embedding multimodal content and users in the joint embedding space. As mentioned in
\autoref{sec:approach}, the proposed model exploits correlations between pairs of multimodal content as well as
content-user pairs.  Due to the disparate nature of different modalities \eg equivalence between content is different in
images and text, and the requirement of evaluating the tasks of both semantic understanding of content and user
interests within the same framework, we have to be extremely careful about creation of training/validation/testing splits. We
proceed by first preprocessing the data in separate steps for textual and visual modalities.

For the textual modality, we first clean the tweets with a basic text processing pipeline that includes removal of
non-alphanumeric characters and stopwords, followed by lemmatization. We use off-the-shelf tools from the StanfordNLP
toolkit for this step \cite{qi2018universal}. We currently retain the hashtags but remove the emoticons embedded in
tweets to simplify our pipeline. In the future, it would be interesting to see the benefits of directly using this
additional information \eg using hashtags as another modality within our framework. We then use the cleaned text to
create a vocabulary (words with at least $5$ occurrences) and learn word-based \emph{word2vec} embeddings
\cite{mikolov2013efficient}. We then use the vocabulary to filter the textual posts for out-of-vocabulary words. In the
multimodal corpora, we found several accounts to be posting content that was either exact duplicates or similar except
for the ordering of one or two words. In order to avoid similar content to be part of both training and test sets, we
remove these duplicates by using a simple procedure based on measuring the percentage of overlapping words between
textual components of two posts along with checking for exact duplicates for the image components. We next pre-process
the visual components of the tweets.

We observe in both corpora that several posted images were exact or near duplicates. This was the case since there could
be re-tweets or a picture could have become viral and different users posted their own interpretations of that image. We
thus begin by creating an index of all the unique images in a corpus. We found standard tools (such as fdupes) based on
image hashes to be ineffective and thus created our own pipeline by using CNN based features for images. To handle a large
number of images, we create a searchable index of all the images in a corpus by using an efficient KD-Tree based
implementation \cite{muja2014scalable}. We then identify nearest neighbors for all the images in the set and iterate over them to create an
index of unique images. We then use this index to create a list of unique image and corresponding captions, that will be
used for creating splits and also evaluating the image-to-text retrieval task \cite{faghri2017vse++}.

Once the modalities have been pre-processed, we now focus on creating splits for evaluation.  The cross-modal retrieval
task demands evaluation of three retrieval tasks- text-to-user, image-to-text, and image-to-user. To perform unbiased
evaluation, we would like to careful about not mixing images/text in the train and validation splits. 
We do so by first creating training/validation/test splits from the visual modality (randomly) by using the list of
(unique) images and captions created earlier. This step ensures that there is no overlap between images between the
train and validation/test splits for evaluating the image-to-text and image-to-user retrieval task. 
In order to create splits for the tasks involving textual modality, we first consider posts with no posted images. 
We then create validation and test sets for evaluating text-to-user retrieval task.
We finally create the training set by merging all the tweets that (i) did not have an image and were not included in the
previous validation/test splits for text-to-user retrieval, and (ii) the tweets with images but not part of the
validation/test splits for retrieval tasks involving images. 
For all the retrieval tasks on the Multimodal Twitter Corpus, we use $20\%$ data for testing,
$2\%$ for model validation, and rest for training. We follow prior works on image-caption matching and use $5000$
image-caption pairs for testing the image-to-text retrieval task \cite{faghri2017vse++}. 
Since our motive for using the Fashion Instagram Corpus is to highlight the effectiveness of the learned user embeddings in
predicting user interests, we maximize the training samples by using only $10000$ posts for the validation set (which is used for
finding the test checkpoint).
The final statistics for both the datasets are
reported in \autoref{tab:corpus}.

\subsection{Evaluation Metrics and Tasks}

We evaluate the joint user and content embeddings learned by our model on two tasks-- (i) cross-modal retrieval, and (ii) prediction
of user interests as a multi-class classification problem. 

We select three cross-modal retrieval tasks to evaluate the learned embeddings for the task of semantic understanding of
content (image-to-caption) and discovering user interest relative to different content modalities (image-to-user retrieval and text-to-user retrieval). Compared to prior works in recommendation based methods that only
evaluate the task of understanding user interests, we
evaluate both the tasks simultaneously.   
Although several metrics exist for evaluating retrieval tasks such as recall, precision,
normalized discounted cumulative mean, we use the mean median rank metric due to its clear interpretation
\cite{manning2010introduction}.  This metric is calculated by first sorting the retrieval results for a query based on
their scores and then computing the median rank of the ground-truth sample. The median ranks across all query examples
are averaged to report the mean median rank.  We use the ground-truth available from the paired content and user
information inside a multimodal post. The best achievable performance for this metric is $1$ when the model always
retrieves the correct result as per the ground-truth.  

We evaluate the task of predicting users interests from the user embeddings learned within our joint embeddings as a
multi-class classification problem for three broad classes-- \quotes{shoes}, \quotes{fashion}, and \quotes{hospitality}.
In this experiment, we aim to investigate the quality of the learned user embeddings for the downstream task of
predicting user behavior.  We annotate the users against the three classes by setting a label for a user and an interest
to $1$ if the user either commented on or followed the brands within that interest group else $0$  (see
\autoref{sec:datasets}).  For this evaluation, we remove the user embeddings for the $23$ initial brands resulting in
$2877$ users. The distribution of the users for each class in shown in \autoref{tab:corpus}.    The experiment is
conducted using 5-fold stratified cross-validation and the reported metric is the F1-score. The F1-score computes the
harmonic mean of the precision and recall from the binary predictions made by the classifier. Since our primary aim is
to establish the quality of the user embeddings for predicting user behavior, we use a simple linear SVM classifier with
$C=1$ for evaluating the embeddings from different models. The performance will generally show the effectiveness of the
user embeddings, learned in an unsupervised manner within our model, in capturing user behavior from their social media
content.

\subsection{Implementation Details}

We use a ResNet-152 network \cite{he2016deep} as the image encoder $\phi_I$. We use the features ($D_I = 2048$) from
a network pre-trained on ImageNet classification task with the last layer removed and do not fine-tune the network during learning. 
We encode a sentence into a stream of vectors by using $300$ word-based embeddings that were learned separately for the
cleaned sentences for each corpus using the Gensim library\footnote{\url{https://radimrehurek.com/gensim/}}. 
We restrict the length of sentences to $20$ words and pad
sentences who length is less than $20$ words with a pad token.
For the text encoder $\phi_T$, we use a CNN, on the word embedding, with three parallel convolutional blocks with different stride lengths ($2,
3, 4$) and number of filter ($512, 256, 256$) \cite{kim2014convolutional}. The outputs from these blocks are
concatenated and used as text features. 
Recently, there has been a significant improvement in pre-trained approaches for several NLP tasks that encode
sentences using deeper and more efficient pipelines \eg BERT and ELMo \cite{devlin2018bert}. However, since our primary
focus was on jointly embedding multimodal content and users in a common geometric space, we opted to keep our
architecture simple and focus more on the core learning method. We set the dimensionality of the common embedding space
as $D=1024$.  We use $300$ dimensional embeddings for users, which are initialized using the average of word2vec
embeddings (learned previously) of all the tweets posted by a user. We found that such initialization helps with faster
convergence. For training our model, we use an Adam optimizer with a learning rate of $0.0005$, which is dropped by a
factor of $10$ after every $10$ epochs. We use a batch size of $1000$ posts and the margin was set to $m=0.2$ in all of our experiments. We select model
checkpoints for evaluation on the test set based on the performance on the validation set. Since we evaluate our model
on $3$ different cross-modal retrieval tasks simultaneously, we use a cumulative metric based on the addition of
normalized mean median ranks on the three tasks for selecting the best checkpoint. We add the normalized ranks for those
retrieval tasks where the regularization parameter is non-zero ($\lambda > 0$).  

\subsection{Quantitative Results}

\subsubsection{Cross-Modal Retrieval}
\label{sec:retrieval}

Since our objective is to simultaneously address the problems of semantic understanding of content and discovering user
interests in a joint framework, we evaluate our model on three cross-modal retrieval tasks on the Multimodal Twitter
Corpus as shown in \autoref{tab:ablation_joint}. We divide the results into three blocks based on the number of modality
pairs being used for training.  Due to the flexibility provided by the proposed loss function,  we are able to control
the relative strength of the different modality pairs\footnote{We refer to users also as a modality in this section.}
for training the joint embedding space by controlling the respective regularization  parameters- $\lambda_1$ for
\pairs{text}{user} pairs, $\lambda_2$ for \pairs{text}{image} pairs and, $\lambda_3$ for \pairs{image}{user} pairs (see
\autoref{eq_loss}). We can also remove the contribution of a modality pair from training by setting the corresponding
$\lambda_i = 0$.  We observe from the results in the first block (\emph{Single Modality Pair}) that the model performs
significantly better than \emph{random} for the modality pair being used for training. For example, for the case of
training with \pairs{text}{user} pair \ie \lambdas{1}{0}{0}, the median rank achieved by \model~on text-to-user
retrieval task is $371$ compared to the random performance of $19904$. Similarly the performance of the model while
training with \pairs{text}{image} and \pairs{image}{user} is $149$ and $1407$ respectively. In the single modality pair
case, the model does not perform well on other modality pairs that have not been used for training.  Although this may
seem obvious, it highlights the disadvantage of prior models (see \autoref{sec:related}) that are either concerned with
content understanding \cite{faghri2017vse++,veit2018separating,datta2019align2ground} or discovering user interests from
a single modality \cite{zhang2019deep,fang2015collaborative}.  As a result, they generally require separate models for
handling each task. 

We now discuss the results reported in the second block (\emph{Two Modality Pairs}), where we use two modality pairs for
training by setting one of the $\lambda_i$ to be always zero. The fact that we are able to control the modality pairs
being used to learn the joint embedding model highlights the generalized nature of our model as compared to prior works
that work with either single or two modality pairs and also do not focus on content understanding.  For simplicity
we do not tune the regularization parameters on the validation set and fix them to $0.5$ for the modality pairs being
used. The median ranks for text-to-user and image-to-text retrieval tasks for the case of \pairs{text}{user} and
\pairs{text}{image} pairs, \ie \lambdas{0.5}{0.5}{0}, are $420$ and $154$ respectively.  Both of these numbers
are slightly lower as compared to their counterparts with single modality pairs ($420$ versus $371$ on image-to-text
retrieval task for the single modality
case).  One can expect such a result since we are handling distant content modalities that are disparate and thus a
dedicated (modality-specific) geometric space might be more optimal for retrieval. 

It is interesting to note that the performance on the image-to-user retrieval ($2455$) tasks is much better than random
($19904$) despite not using that modality pair for training.  We believe this happens since the joint embedding space is
indirectly able to associate images and users by learning to correlate \pairs{text}{user} pairs along with
\pairs{image}{text} pairs. The textual space is able to act as a bridge between users and images.  This is a significant
outcome since  it highlights the ability of our model to discover joint representations that can reason about a modality
pair without ever having seen it during training. This is also an example of zero-shot retrieval \cite{bansal2018zero}
\cite{fu2017recent,dalton2013zero}, where we are able to retrieve samples from unseen modality pair(s) during testing. 
Interestingly, we also make a similar observation for the case of training
with \emph{(text, user)} and \emph{(image, user)} modality pairs \lambdas{0.5}{0}{0.5}. The
performance on the image-to-text retrieval task, for which no modality pairs have been observed during training, is $239$,
which is noticeably better than random performance of $2500$. This result highlights that users are not merely a
separate or disconnected entity in a social media network in comparison to the content. Rather, they seem to be acting as
anchors (or topics) in the embedding space that enhance the semantic understanding of multimodal content in this space.
This is a promising result for future research on learning user behavior and content semantics in social media network
since it shows that despite the unconstrained nature of the content and user subscription, the data is in fact quite structured. As a result, the
joint understanding of content and users can be used to improve parallel research on multimodal content understanding.

We also note in the two modality pair case that for $\lambda_3 \neq 0$, the model performs quite well on the
image-to-user retrieval task as compared to using only \emph{(image, user)} pairs for training. For example, the
median ranks for \lambdas{0.5}{0}{0.5} and \lambdas{0}{0.5}{0.5} are $664$ and $1181$ respectively as compared to $1407$
of \lambdas{0}{0}{1}. We believe this happens because the image-to-user retrieval task is generally harder as compared
to text-to-user retrieval owing to the complexity of the visual modality and also because the training data is smaller as
compared to the latter case (the ratio for volume of image to text content is approximately $1:10$). In this case, being able
to utilize correlation information from other modality pairs is helpful, as has been observed in prior works on
multi-task learning with several multimodal tasks \cite{kaiser2017one}, where information from additional modalities is
able to regularize tasks with limited data or higher complexity.  Such multi-task learning is more useful when learning
along with \pairs{text}{user} pairs since the task is directly related to understanding user interests and retrieving
them based on the content. Interestingly, we also note that learning to associate content from multimodal content pairs \ie 
\pairs{image}{text} pairs, also aids in discovering user interests.

We show results for training the model with all the modality pairs in the third block (\emph{All Modality Pairs}) in
\autoref{tab:ablation_joint}. We show results with three different combinations of the regularization parameters. The
first combination \lambdas{0.33}{0.33}{0.33} gives equal weights to all the regularization parameters. While the second
and third combination give a lower weight to $\lambda_1$, which corresponds to \pairs{text}{user} modality pair- which
has a larger number of samples compared to modality pairs involving images, the third combination
\lambdas{0.1}{0.45}{0.45} performs best on all the tasks based on a combined metric that computes the sum of normalized
median ranks for the three tasks. First, we observe for \lambdas{0.33}{0.33}{0.33} that our model performs quite well
on all the cross-modal retrieval tasks. For example, the performance on user-to-text, image-to-text, image-to-user is
$450$, $144$, and $980$ respectively. Although these numbers are slightly lower as compared to the best performance
recorded in the previous case, the performance is still quite competitive given that we are using joint representations
from a single multimodal space to tackle three retrieval tasks with distant modalities. To the best of our knowledge,
such an observation has not been shown in any prior work working with multiple domains \cite{elkahky2015multi,zhang2017joint}. 
While training our model with the third combination (\lambdas{0.1}{0.45}{0.45}), the model performs best on the
image-to-text retrieval task ($127$ compared to previous best of $138$ for \lambdas{0}{0.5}{0.5}). We also notice the
performances on text-to-user and image-to-user retrieval tasks are at par or better compared to the case of training the
model with individual modality pairs. For example, the performance on image-to-user with training using all modality
pairs (\lambdas{0.1}{0.45}{0.45}) and only \pairs{Image}{User} pairs (\lambdas{0}{0}{1}) is $785$ and $1407$ respectively.
This highlights the strength and advantages of our model \model, which learns to jointly embed multimodal content and
users in the common embedding space.

These findings are important to the research community since they clearly  show that content understanding and discovering
user behavior from social multimedia data can be tied together in a single optimization problem. This is possible due to
the underlying structure of users, their behavior, and the posted content on social media networks. We believe that we
observe better performance on content understanding (image-to-text) while using information from users' textual and
visual posts since the users serve as anchor points, in the embedding space, that are connected with specific content
semantics. Such anchors act as regularizers to improve performance on tasks with limited training data. We also note
the performance improvement on the image-to-user retrieval task while using information from either \pairs{image}{text}
pairs or \pairs{text}{user} pairs or both for training. The proposed model is a generalization of prior works that
either use single modality pairs for training or train with multiple domains from a single modality (generally textual)
without associating content from multiple domains. We argue that this work will help merge content understanding with
modeling user behavior for future research. Moreover, the proposed model is general enough to handle other modalities
such as speech, hashtags, emoticons.

\subsubsection{Predicting User Interests}
\label{sec:user_interests}

The results presented in the previous section show that our model is able to perform well on retrieval tasks related to
both content understanding and discovering user interests. We previously evaluated the task of understanding user
interest by using two cross-modal retrieval tasks to retrieve users from either text or images. We now
try to ascertain the quality of the learned user embeddings directly for predicting user behavior/interests. The
experiments are conducted on the Fashion Twitter Corpus and the results are presented in  
in \autoref{tab:user_behavior}. We first report results with two baselines where the user
embeddings are obtained by averaging the word embeddings of their posts (\emph{Avg-Text}) and the image embeddings of
their posted images (\emph{Avg-Image}). We observe that the image based embeddings (F1-score=$0.40$) outperform the text
based embeddings (F1-score=$0.45$). We believe this happens because these experiments are conducted on a dataset
collected from Instagram, which is a highly visual social network. As a result, images are
generally more representative of a user behavior as compared to their posted text. We also show results with the
proposed model for different combinations of modality pairs being used for training. These combinations (in sequence)
use \pairs{text}{user} pairs, \pairs{image}{user}, and all content/user pairs for training the joint embeddings within
the proposed framework (see \autoref{sec:retrieval}). We use \lambdas{0.33}{0.33}{0.33} for training the model with all
content/user pairs since the ratio of text to images is almost the same for this dataset.    

The general performance trends seem to be similar to those presented in the cross-modal retrieval experiments. The model
using \pairs{text}{user} pairs for training seems to perform worse as compared to the model using \pairs{image}{user}
pairs (F1-score of $0.47$ vs. $0.49$). This trend is expected due to the visual nature of Instagram posts as also seen
for the baseline models.  We also observe the advantage of learning these user embeddings within our model instead of
naively computing them by averaging, as done for the baseline. For example, our model using \pairs{image}{user} pairs has a
F1-score of $0.49$ compared to $0.45$ of the baseline model Avg-Image that also uses image posts from the users.  The
proposed model while using all the modality pairs for learning achieves the best performance on this task
(F1-score=$0.52$ compared to previous best of $0.49$ for \pairs{image}{user}\footnote{The result is also statistically
significant with a p-value $<0.05$.}). This result highlights the fact the joint understanding of content along with user
behavior allows our model to learn stronger representations for each user that effectively capture their underlying
behavior. Being able to learn user embeddings that are informed by multimodal content understanding seems to provide an
alternative and can also be used in combination with methods utilizing information about social connections such as followers, friends,
\cite{grover2016node2vec,liao2018attributed,lee2011understanding} for improving the understanding of user
behavior.

\begin{table*}
	\centering
	\begin{tabular}{ccc|c||ccc}
		\hline
		\multicolumn{3}{c|}{Modality pairs used for training} & Regularization & \multicolumn{3}{c}{Mean Median Rank} \\
		 &  \model &  & Parameters &   & Cross-Modal Retrieval  &   \\ 
		 &  &  & $(\lambda_1, \lambda_2, \lambda_3)$ &   &  &  \\ \hline
		Text-User & Image-Text & Image-User & & Text-To-User & Image-To-Text & Image-To-User \\ \hline
		\multicolumn{3}{c}{\emph{Random}} & & $19904$  & $2500$  & $19904$   \\ \hline 
		\multicolumn{6}{c}{Single Modality Pair} &  \\ \hline
		\checkmark &  &  &  $(1, 0, 0)$ & $\textbf{371}$  & $2291$  & $21873$   \\ 
		&  \checkmark  &  & $(0, 1, 0)$ & $20212$  & $149$  & $20289$   \\ 
		&  &   \checkmark & $(0, 0, 1)$ & $19698$  & $2491$  & $1407$   \\  \hline
		\multicolumn{6}{c}{Two Modality Pairs} &  \\ \hline
		\checkmark&  \checkmark&  &  $(0.5, 0.5, 0)$ & $420$  & $154$  & $2455$   \\ 
	       \checkmark       &  & \checkmark &  $(0.5, 0, 0.5)$ & $383$  & $239$  & $\textbf{664}$   \\ 
		&  \checkmark  & \checkmark &  $(0, 0.5, 0.5)$ & $8427$  & $138$  & $1181$   \\  \hline
		\multicolumn{6}{c}{All Modality Pairs} &  \\ \hline
		\checkmark &  \checkmark  & \checkmark &  $(0.333, 0.333, 0.333)$ & $450$  & $144$  & $980$   \\ 
		\checkmark &  \checkmark  & \checkmark &  $(0.05, 0.2, 0.75)$ & $732$  & $140$  & $806$   \\ 
		\checkmark &  \checkmark  & \checkmark &  $(0.1, 0.45, 0.45)^*$ & $551$  & $\textbf{127}$  & $785$ \\

	\end{tabular}

	\caption{Table shows the performance on cross-modal retrieval tasks when training the proposed model with a
		single modality pair, two modality pairs, and all modality pairs on the Multimodal Twitter Corpus. For each row, we also show the
		regularization parameters used for the proposed loss function (\autoref{eq_loss}) that control the
		contribution of different modality pairs for training our model.  
		The cross-modal retrieval task measures the performance for both the tasks of content understanding (Image-To-Text) and
		discovering user interests (Text-To-User and Image-To-User).
		$^*$:
		Denotes the model that performs cumulatively best on all the tasks based on the sum of normalized mean
	median ranks for the three tasks (see \autoref{sec:retrieval}). }   

	\label{tab:ablation_joint}
\end{table*}

\begin{table}[htbp]
	\centering
		\begin{tabular}{c|c|c|c} 
			\hline
			Method & \multicolumn{2}{c|}{Modalities used} & F1-score \\
			& Text & Image  &   \\ \hline
			Random & & & $0.00$ \\ \hline \hline
			\multicolumn{4}{c}{Baseline} \\ \hline
			Avg-Text & \checkmark & & $0.40$ \\ \hline
			Avg-Image &  & \checkmark & $0.45$ \\ \hline \hline
			\multicolumn{4}{c}{\model~(Proposed)} \\ \hline 
			Text-User pair & \checkmark & & $0.47$ \\ \hline
			Image-User pair & & \checkmark & $0.49$ \\ \hline
			All modality pairs & \checkmark & \checkmark & $\textbf{0.52}$ \\ \hline
		\end{tabular}

		\caption{Table showing results on the downstream task of predicting user interests from the user
		embeddings learned by our model on the Fashion Instagram Corpus (see \autoref{sec:user_interests}).}

		\label{tab:user_behavior}
\end{table}

\subsection{Qualitative Results}

\subsubsection{Discovered Multimodal User Clusters}
\label{sec:clusters}

\begin{figure*}[htbp!] 
	\centering
	\begin{center} 		
		\includegraphics[width=0.7\linewidth]{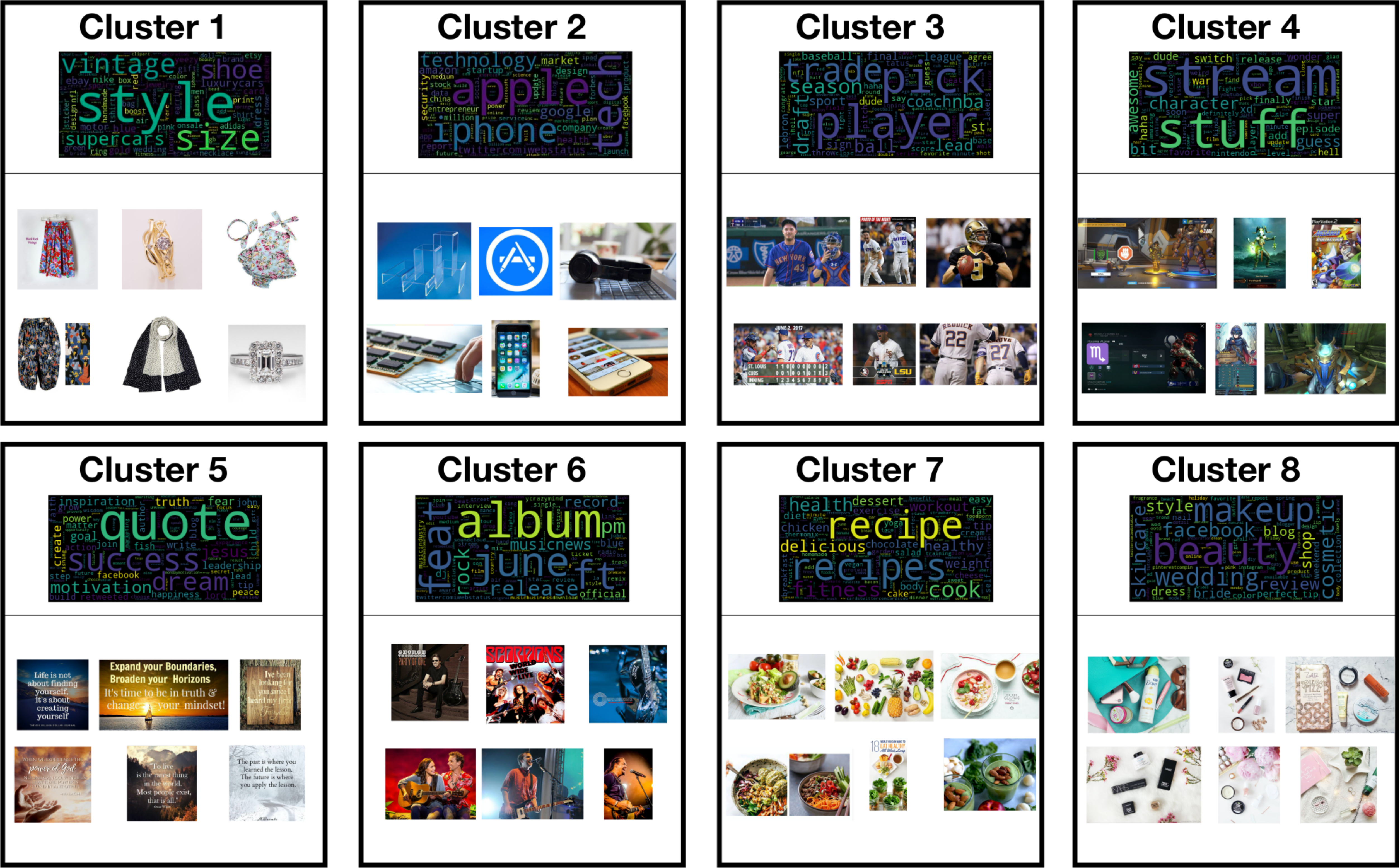}
	\end{center} 

	\caption{This figure shows the multimodal user clusters discovered by our model. For each user cluster we also show 
	the wordcloud for the top words corresponding to the users in that cluster as well as the images closest
to the centroid of these clusters in the common embedding space (see \autoref{sec:clusters}).}

	\label{fig:cluster} 
\end{figure*}

In order to gain further insight into the learned user embeddings and how they relate to multimodal content, we analyze
them using unsupervised clustering. To do so, we first cluster the user embeddings learned for all the users from the
Multimodal Twitter Corpus using k-means into $50$ clusters. We use the embeddings from the best performing model in
\autoref{tab:ablation_joint} using all the modality pairs (\lambdas{0.1}{0.45}{0.45}). In order to ground these clusters
within the visual and textual social multimedia content, we create a frequency table of all the words present in the
posts corresponding to the users within each cluster.  From these word-frequency tables, we remove the most common $200$
words in the corpus. The word-frequency tables are then used to create wordclouds, where the size of each word is
dependent on its frequency. In order to select the representative images for each cluster, we use the property that
images and cluster centroids-- calculated as average of the embeddings of the users in that cluster, are embedded in the same space.
We select the top images based on their proximity to the cluster centroids in the embedding space. We have shown the wordclouds and top images
for six clusters in \autoref{fig:cluster}.

These clusters clearly show that the clusters emerging from the learned user embeddings are naturally grounded in the multimodal content. The learned clusters
seem to be generally pure and represent semantically meaningful topics. For example, cluster-1 seems to correspond to
fashion, shoes \etc, while cluster-2 corresponds to technology, gadgets \etc. It is also interesting to note that our
model is also able to discover slightly abstract topics such as cluster-5 which corresponds to motivational post quotes.
We also observe two clusters (1 and 8) with semantically close concepts related to fashion. Cluster-1 seems to represent to
vintage fashion, clothing \etc, while cluster-8 represents cosmetic products. Being able to discriminate between
semantically close concepts shows the ability of our model in being able to successfully discover semantics from in-the-wild
social multimodal data.  These visualizations highlight the strength of our model in being able to not only establish
content understanding but also discover user interests grounded in those semantics. We believe that such an
understanding also provides a powerful tool to index and search large-scale social multimedia data for different topics.

\subsubsection{Semantic Cross-Modal Retrieval Examples}
\label{sec:ret_examples}

\begin{figure*}[htbp!] 
	\centering
	\begin{center} 		
		\includegraphics[width=0.6\linewidth]{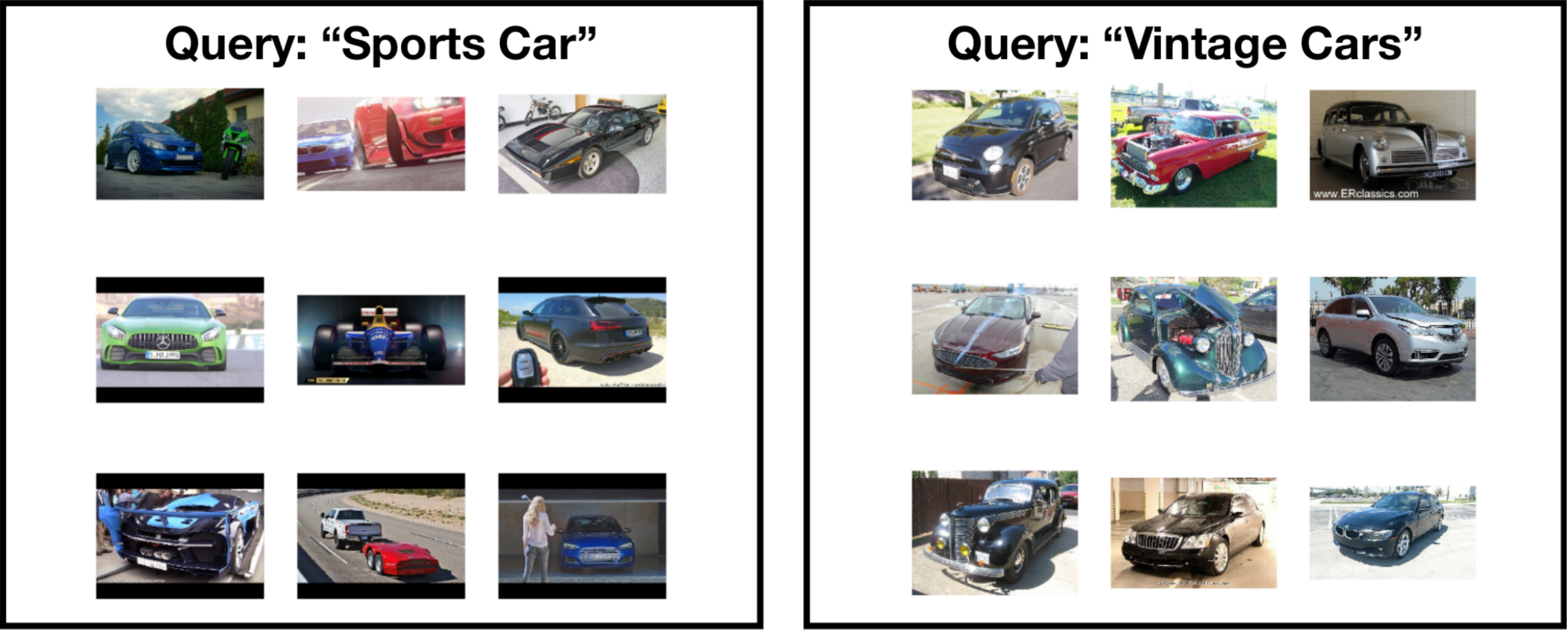}
		\includegraphics[width=0.6\linewidth]{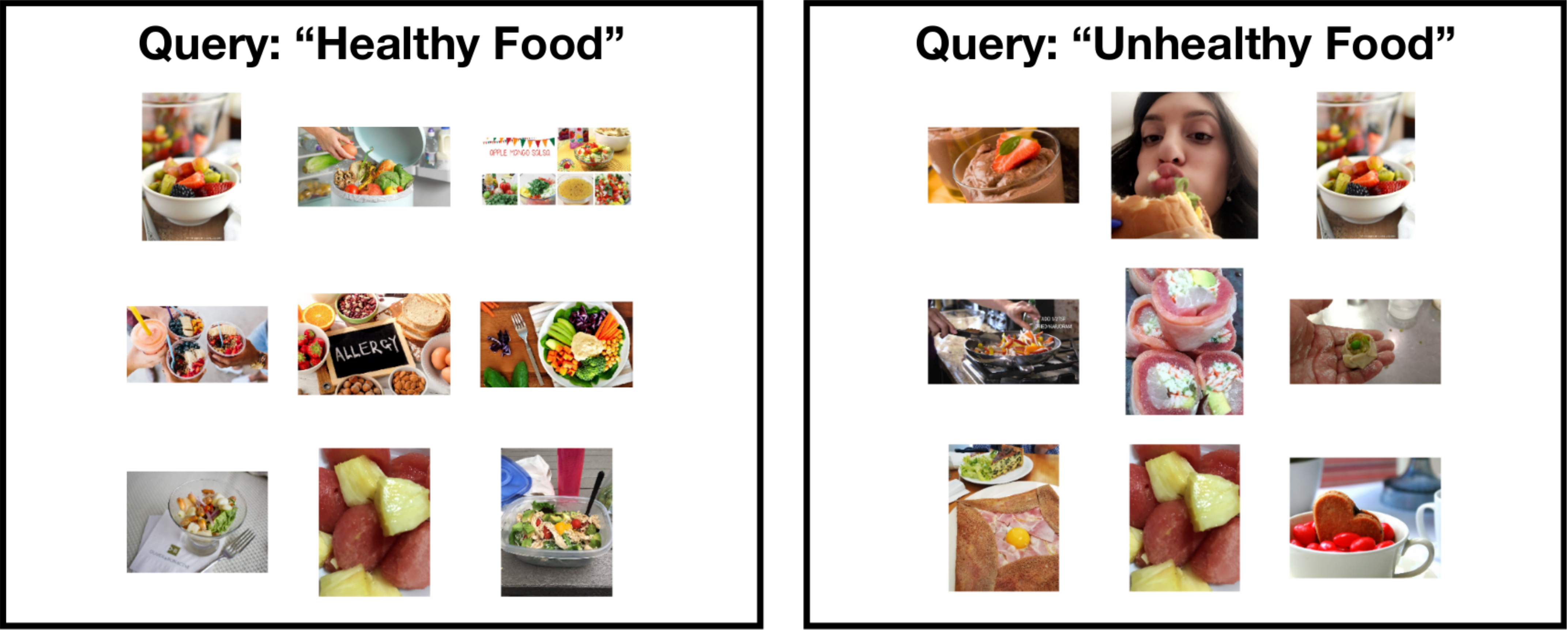}
		\includegraphics[width=0.6\linewidth]{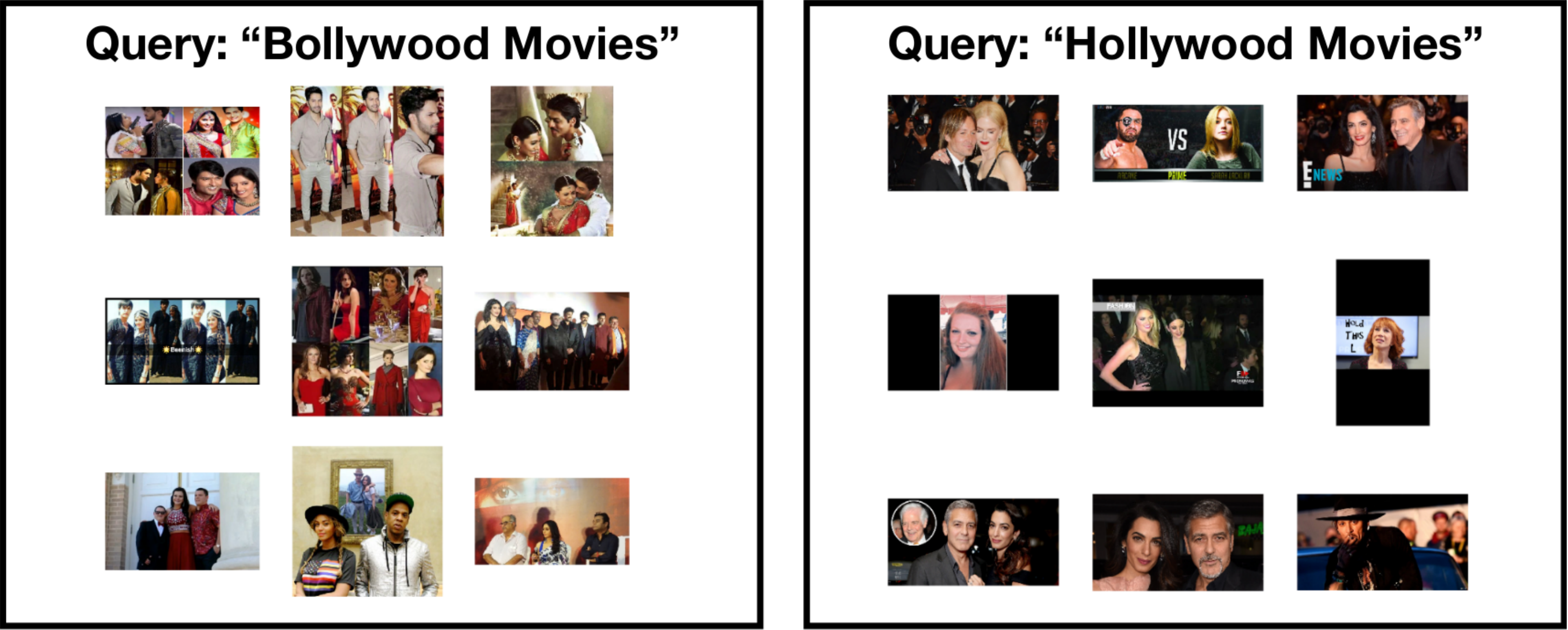}
	\end{center} 
	\caption{This figure shows the top images retrieved for two semantically closely queries using the multimodal
	embedding learned by our model. This highlights the ability of our model to semantically understand multimodal
content from noisy data from social media networks (see \autoref{sec:retrieval}).}
	\label{fig:retrieval} 
\end{figure*}

We now try to gain insights into the ability of our model to understand multimodal content. We refer to content
understanding as the ability to automatically associate high-level or semantic meaning described using textual modality
with visual modality (see \autoref{sec:related}). We show top retrieved images for few closely related textual queries
in \autoref{fig:retrieval} for the model learned using all the modality pairs (\lambdas{0.1}{0.45}{0.45}) on the
Multimodal Twitter Corpus. We observe that our model is able to effectively discriminate between semantically close
concepts such as \quotes{sports car} and \quotes{vintage cars} or \quotes{healthy food} and \quotes{unhealthy food}.
Such a result shows the strength of our model in being able to understand highly unstructured multimodal content from
social media data. At the same time, it also shows the advantage over prior works utilizing discrete hashtags
for describing images due to their restriction by the given hashtag vocabulary \cite{veit2018separating}. On the other
hand, the proposed model is able to utilize the compositional nature of language to discover new meanings in the visual
modality. We believe that this will encourage further research on learning to describe content by learning from
weakly/webly supervised and massive social multimedia data. As shown in \autoref{tab:ablation_joint}, the ability to understand
content is also enhanced by joint learning of user embeddings which seem to act as anchors that act as regularizers. This shows
that content and user behavior are well connected in social multimedia networks.

\subsection{Comparison with Competing Methods}
\label{sec:sota}

We compare our model on the tasks of cross-modal retrieval with a simple baseline and two competing approaches using
the Multimodal Twitter Corpus in \autoref{tab:comparison}.  In this experiment, we want to establish that the proposed
model is able to perform well together on the tasks of user behavior discovery and understanding user content.  The
former task is evaluated using content-to-user retrieval tasks (image-to-user and text-to-user), while the latter task is
evaluated using content-to-content retrieval tasks(image-to-text), as described in \autoref{sec:retrieval}.  The combined
performance on all these tasks is computed by adding their normalized (using the total number of retrievable samples) mean median ranks.  We first implement two
baseline approaches-- \emph{Avg-Text} and \emph{Avg-Image}, where the user embeddings are computed by averaging the word
embedding of their tweets and the image embeddings of the posted images respectively. We use a random vector for users
having no image content.  Since these methods only yield embeddings for content-user pairs in their respective cases,
their performance can only be computed for text-to-user and image-to-user retrieval tasks.  The other two approaches are based
on prior recommendation based methods that either learn from a single modality \cite{he2016deep,fang2015collaborative} or
multiple modalities without any understanding of content \cite{elkahky2015multi,zhang2017joint}.  The first approach is
based on extending previous recommendation methods that learn joint embeddings for single modality-user pairs to handle
content understanding between multiple modalities along with user behavior. We refer to this approach as
\emph{Bridging-Modality}, where the idea is to first learn content-user embedding space for a single modality and then
use that modality as a bridge to connect the other modality and the user embeddings.  We achieve this by training our
method first with only \pairs{text}{user} pairs (\lambdas{1}{0}{0}). Thereafter, we fix the encoders for the text and
the user modality and learn the model with \pairs{image}{text} pairs (\lambdas{0}{1}{0}).  Since we had earlier
constrained the joint embedding space by fixing the networks for user and text modalities, we are now able to associate
images and users implicitly by using the text modality as a bridge to connect them. The second approach, referred to as
\emph{Merged-Modality} is based on \cite{elkahky2015multi,zhang2017joint}, where the features from the two modalities
are merged together and a single loss function is using for embedding the merged content and users in a common space. In
this case, we merged the image and textual content by fusing them with sum pooling after they are embedded in the same
space. We handle the case of missing data for image modality by using an embedding vector filled with zeros. We allow a
fair comparison by using the same learning settings and ranking loss formation as used in our model.

The results in \autoref{tab:comparison} reveal that the proposed model \model~outperforms the baseline methods and the
methods based on prior state-of-the-art methods on the combined task of understanding content and discovering user
behavior by a noticeable margin. Our model is able to surpass the baseline on both the image-to-text retrieval tasks
($1372$ of baseline versus $785$ of \model) and text-to-user retrieval task ($1963$ of baseline versus $551$ of \model).
This highlights the benefits of using learning based paradigms for learning recommendation models as compared to static
approaches.  We also observe that our model outperforms the Bridging-Modality based method on the image-to-text ($238$
of Bridging-Modality versus $127$ of \model) and the image-to-user ($2083$ of Bridging-Modality versus $785$ of \model)
retrieval tasks by a non-trivial margin. Although the performance of our model on the text-to-user retrieval ($551$)
task is slightly lower as compared to Bridging-Modality ($372$), it outperforms the latter method on the normalized mean
rank metric\footnote{The metric is normalized mean median rank and thus lower value is better.}measuring the joint
performance on all the tasks ($0.042$ of \model~versus $0.062$ of Bridging-Modality). This result highlights the ability
of our model to effectively handle heterogeneous content from social media networks that can occur in different
proportions due to the specific nature of the social media network. For example, networks such as Twitter and Reddit are
more text-focused compared to Instagram.  Moreover, the performance improvements on the image-to-user retrieval task
demonstrates that Bridging-Modality based model is unable to effectively utilize the relationship between content and
user interests to enhance the performance on both the tasks, as achieved within our model (see \autoref{sec:retrieval}).
We also observe that our model performs significantly better than does the Merged-Modality based method, which merges different
modalities together and uses a single loss function for correlating content and users \cite{zhang2017joint}. Both of
these methods-- Bridging-Modality and Merged-Modality, seem to be doing well on the text-to-user retrieval tasks, which
has more data, but perform poorly on image related retrieval tasks.  These results demonstrate the strength of our model
in tackling a key limitation in prior works that focused solely on discovering user interests and did not address
content understanding in a holistic manner. A key insight in these results is that social media networks, despite their
unconstrained nature, are a structured ecosystem where content and users interact together to produce meaning.

\begin{table*}
\centering
\begin{tabular}{c||cccc} \hline
                 Method & \multicolumn{4}{c}{Mean Median Rank} \\ \hline
			&  Text-User & Image-Text & Image-User & Joint-Performance \\ \hline
		 Random & $19904$ & $2500$ & $19904$ & --   \\ \hline			
		 Baseline (Avg-Text) & $1963$  & -- & -- & --  \\ \hline
		 Baseline (Avg-Img) & -- & -- & $1372$ & --  \\ \hline
		 Bridging-Modality & $\textbf{372}$ & $238$ & $2083$ & $0.062$  \\ \hline
		 Merged-Modality & $514$ & $601$ & $2405$ & $0.148$  \\ \hline
		 \model~(Proposed) & $551$ & $\textbf{127}$ & $\textbf{785}$ & $\textbf{0.042}$  \\ \hline			
                 
\end{tabular}
\caption{Comparison of proposed model with different competing approaches on cross-modal retrieval task on the
	Multimodal Twitter
Corpus. The joint performance is computed by adding the normalized (using the total number of retrievable samples) mean median ranks across the three tasks.}
\label{tab:comparison}
\end{table*}

\section{Conclusion and Future Work}

We presented a novel content-independent content-user-reaction model for social multimedia that embeds  users, images
and text drawn from open social media in a common multimodal geometric space, thereby enabling seamless multi-way
retrieval.  Our approach is able to simultaneously tackle the two problems of semantic understanding of multimodal
content and modeling user interest within a unified mathematical framework. Our model uses a novel loss function that
allows us to handle the complexities associated with the unconstrained nature of social multimedia data.  We established
the validity of our approach by achieving consistent improvements beyond prior art on several cross-modal retrieval
tasks on a real-world multimodal dataset collected from Twitter. We showed that our model is able to exploit the
relationships between content and user behavior on social media networks to not only improve the performance on
retrieval tasks for modalities with limited data but also perform zero-shot retrieval. We applied user embeddings
learned
from our joint multimodal embedding to the task of predicting user interests in an Instagram based dataset, and showed
that the best results are obtained by the joint image-text-user embedding compared to all other combinations of
modalities. Since our solution is based on learning a similarity measure between modalities, it yields a general purpose framework for finding user
affiliations without the use of linking data from the social network. Our framework is able to deal with the inherent lack
of explicit structure and grammar and unconstrained subject matter of social media.

Our results showed that the best retrieval performance on the combined tasks of content understanding and user interest
discovery is achieved when all modalities are used which in turn shows that
social media content is inherently multimodal even when a social network lends itself best to a single modality, such as
text for Twitter and images for Instagram.  Despite the lack of explicit structure in social network membership and
content, our work reveals emergent structures in content-user relationships, as well as in content-content relationships
within and across modalities. The clusters of content that emerge are remarkably consistent in their semantics. Our
framework sets up the possibility of systematically understanding the relative significance of each modality in exerting
influence. 
 
The proposed framework is highly scalable and also lends itself to seamless incorporation of additional modalities. We
have some early results with embedding sentiment and document intent that indicate that we can get a richer
understanding of the influence exerted by a piece of multimedia content than we do currently. The proposed three way
retrieval sets up the possibility of generating content to better suit the interests of a certain group by enabling
modality wise exploration and retrieval of more suitable content, as well as achieve better coherence in the generated
content. Moreover, our framework can be combined with methods that explicitly use social network graph for learning user
representation. Finally, our multimodal embeddings based features are general purpose since they can be used to drive a
variety of event detectors on other downstream tasks. Our framework thus offers rich possibilities for further work in
social media content understanding and generation.

\section{Acknowledgment}

This project is supported by the USAMRAA under Contract No. W81XWH-17-C-0083.
\textbf{Disclaimer:} The views, opinions, and/or findings expressed are those
of the author(s) and should not be interpreted as representing the official
views or policies of the Department of Defense or the U.S. Government.

We would like to thank (1) Jonathan Pfautz, Laurie B. Weisel and Brian Dennis from DARPA for
helpful discussions, (2) Indranil Sur for help with collection of the Instagram
corpus, and (3) Vadim Kagan, Skylar Durst, and VS Subrahmanian for helpful
discussions. We would also like to thank our interns and co-workers- Suhas
Lohit, Ankan Bansal, Julia Kruk, Carter Brown, Karuna Ahuja- whose work
contributed indirectly towards the development of this work. We would also like to thank Mohamed Amer for informing us about this venue.  


{\small
\bibliographystyle{spmpsci}
\bibliography{egbib}
}

\end{document}